\documentclass[prl,aps,twocolumn,showpacs]{revtex4}
\usepackage{graphicx}

\newcommand{\be}{\begin{equation}}
\newcommand{\ee}{\end{equation}}
\newcommand{\beqa}{\begin{eqnarray}}
\newcommand{\eeqa}{\end{eqnarray}}

\begin{document}

  \title{Nonsuperfluid origin of the nonclassical rotational inertia in a bulk sample of solid $^4$He}

\author{John D. Reppy$^\ast$}
\affiliation{Laboratory of Atomic and Solid State Physics, Cornell University, Ithaca, New York 14853-2501}

%\date{March 26, 2010}

\pacs{67:80.bd,66.30.Ma}

\begin{abstract}The torsional oscillator experiments described here examine the effect of disorder on the nonclassical rotational inertia (NCRI) of a solid $^4$He sample. The NCRI increases with increasing disorder, but the period changes responsible for this increase occur primarily at higher temperatures. Contrary to expectations based on a supersolid scenario, the oscillator period remains relatively unaffected at the lowest temperatures. This result points to a nonsuperfluid origin for the NCRI. 
\end{abstract}

\maketitle

The work reported here investigates the changes in the period of a torsional oscillator (TO) containing a solid hcp $^4$He sample as a result of plastic deformation of the sample. In the pioneering experiments of Kim and Chan$^{1,2}$ (KC) at Penn State, a decrease in the period of a TO containing a solid $^4$He sample was observed for temperatures below 250 mK. KC interpreted the period decrease in terms of an onset of the growth of non-classical rotational inertia (NCRI)$^3$ arising from a superfluid-like decoupling of a fraction of the solid $^4$He moment of inertia from the TO.  In subsequent work, Kim et al$^4$ showed that the magnitude of the non-classical rotational inertia fraction (NCRIF) is sensitive to $^3$He impurity concentrations. The Penn State group also found a sensitivity to the method of sample preparation.$^5$ The experiments of Rittner and Reppy$^6$ demonstrated that annealing of the solid $^4$He samples can greatly reduce the magnitude of the NCRI signal, suggesting that the level of disorder in the sample plays an important role in this phenomenon.

In a series of interesting experiments, Day and Beamish$^7$ (DB) have studied the shear modulus, $\mu$, of solid $^4$He.  They find an anomalous increase, $\Delta \mu$, in the shear modulus at low temperatures which bears a striking resemblance to the temperature dependence of the NCRI signals observed in the TO experiments. DB argue that the temperature dependence of the shear modulus and the related phenomena can be understood in terms of the pinning of dislocation lines by $^3$He as the temperature is lowered. Although it is tempting to attempt to explain the TO results in terms of the changes in the shear modulus, calculation$^8$ shows that the 10 to 20\% variations seen in the modulus by DB are far too small to explain the TO results. West et al$^9$ also report an increase in the shear modulus for hcp $^3$He. However, the NCRI signature is absent in $^3$He experiments, which suggests a role for quantum statistics in these experiments.

The aim of the current experiment was to determine the influence on the NCRI of disorder induced by the plastic deformation of a solid $^4$He TO sample. Plastic deformation of solid $^4$He is known to result in the creation of dense populations of dislocations in the solid.$^{10-13}$ The deformation of the solid and the accompanying generation of a high-density population of dislocations was expected to lead to an increase in the NCRI signal, as has been confirmed by this experiment.

The TO cell used for the current experiments is shown in Figure 1.  At low temperatures the frequency of the TO is 818 Hz and the rim velocity is in the range of 10 to 20 $\mu$m/s.    An inner cylinder, attached to the flexible diaphragm at the top of the cell, defines an annular region containing the solid $^4$He sample. It has a height of 0.917cm, a radial gap of 0.032 cm, and a mean radius of 0.714 cm.  The gaps at the top and bottom of the inner cylinder are 0.025 cm.  The TO period at liquid helium temperatures is 1.219 ms and the total period shift for a solid sample with a density of 0.200 gm/cm$^3$ is calculated to be 185 ns.   A Straty-Adams (SA) capacitance gauge$^{14}$ with mbar pressure sensitivity is mounted on the upper side of the diaphragm and the region surrounding the gauge is enclosed, with a fill line added to allow the volume above the diaphragm to be filled with liquid $^4$He.  Hydrostatic pressure can then be applied to the diaphragm to produce a controlled displacement of the diaphragm and the attached inner cylinder. More details of this technique are given in Rittner et al.$^{15}$ The displacement is monitored by the capacitance of the SA gauge, suitably corrected for the pressure-dependent dielectric constant of the liquid helium between the plate capacitor plates. Since the TO has a pressure sensitivity of 0.88 ns/bar, the sample pressure is monitored with the SA gauge to insure that period shifts observed in the course of the experiment are not due to pressure changes within the cell.

\begin{figure}[htbp]
\begin{center}
\includegraphics[width=0.42\textwidth]{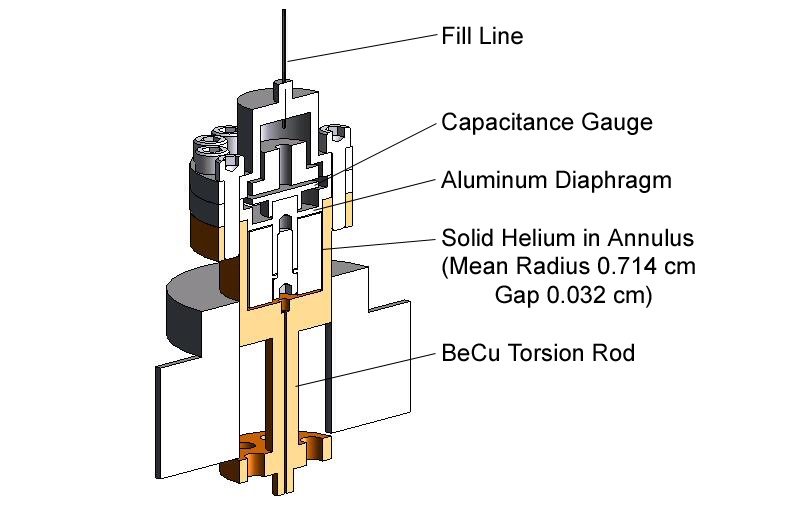}
\caption{The torsional oscillator, consisting of three sections, is shown in cross section.  The body and torsion rod are made from a single piece of hardened BeCu alloy.  The other sections, including the flexible diaphragm, were machined from a high strength aluminum alloy.  The Straty-Adams gauge electrodes are brass, and the upper fill line consists of a section of CuNi tubing with an internal diameter of 0.01 cm.}
\label{default}
\end{center}
\end{figure}

The maximum pressure that can be applied to the diaphragm is limited to the melting pressure of $^4$He at low temperature. At this pressure, the inner cylinder is displaced by 1.9 $\mu$m relative to the outer wall of the cell, resulting in a shear strain in the annulus of 5.9 x 10$^{-3}$.  This level of strain is well above the threshold for the plastic deformation of hcp $^4$He.$^7$ The solid $^4$He samples are formed by the blocked capillary (BC) method, using standard $^4$He (0.3 ppm $^3$He impurity).  The cell is initially filled with liquid to a pressure near 60 bar, then cooled to freeze the sample. At the completion of  freezing the sample pressure has been reduced to about  40 bar.  Sample formation by the BC technique results in a polycrystalline sample with a relatively high degree of initial disorder.  
\begin{figure}[htbp]
\begin{center}
\includegraphics[width=0.42\textwidth]{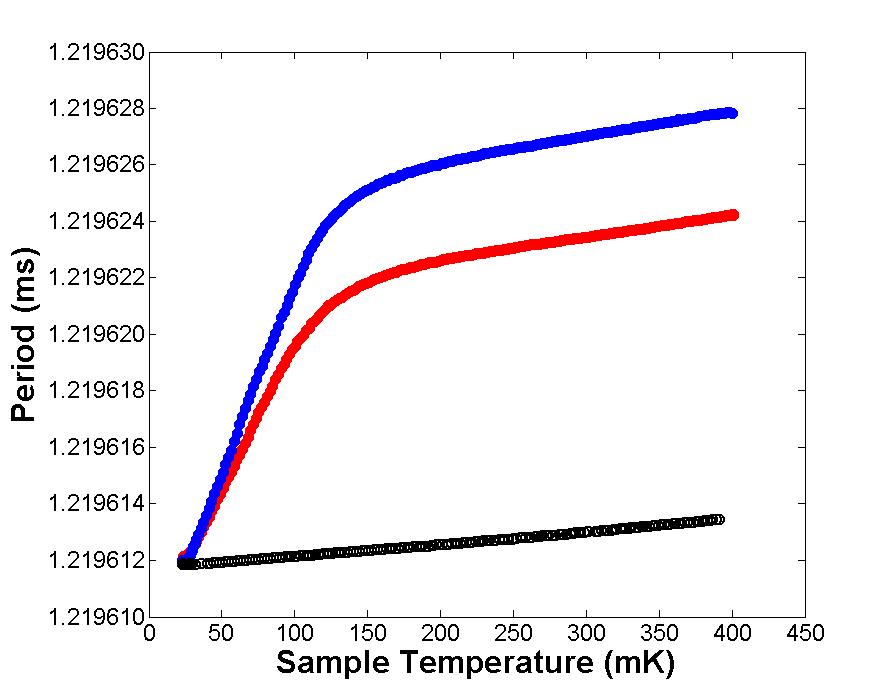}
\caption{Period data for the initial sample of bulk solid $^4$He are shown in red as a function of temperature.  The period data obtained following the deformation of the sample by an application of 10.3 bar to the flexible diaphragm are plotted in blue.  The bottom data, plotted in black, are the period data for the empty cell shifted upward by 110 ns.}
\label{default}
\end{center}
\end{figure}
In an attempt to reduce the initial degree of disorder, most samples are annealed for up to 20 hours at temperatures above 1 K.  In Figure 2 the period data are shown for a $^4$He sample with two different levels of disorder. Period data for the empty cell, shifted to match the other data sets at 20 mK, are also shown. The magnitude of the NCRI signal, is determined in the usual way by subtracting the raw period data from a linear ``background'' function determined by a fit to the period data above the onset temperature.  The initial sample, middle data set in Figure 2, was cooled to 20 mK after being annealed at 1.6 K.  The magnitude of the NCRI for this sample was 9.1 ns for a 4.9\% NCRIF . The temperature was raised then to 250 mK, and $^4$He was slowly condensed into the volume above the diaphragm. When this volume was full, the pressure was raised to 10.3 bar, to produce a 0.81 $\mu$m displacement of the inner cylinder and a shear strain of 2.55 x 10$^{-3}$. The pressure was then reduced to near zero by pumping on the fill line, thus returning the diaphragm and inner cylinder to their original positions. The upper data set was obtained following the deformation of the sample. The NCRI has now increased to 12.6 ns or a 6.8\% NCRIF. There are two important features: 1) the TO period is increased over a wide temperature range above the ``supersolid" onset; 2) the increase in the period approaches zero as the temperature is reduced to 20 mK. 

The effect of the deformation on the TO period is most clearly displayed by subtracting the period data for the undeformed sample from the period values after deformation. These data are shown in Figure 3. 
\begin{figure}[htbp]
\begin{center}
\includegraphics[width=0.42\textwidth]{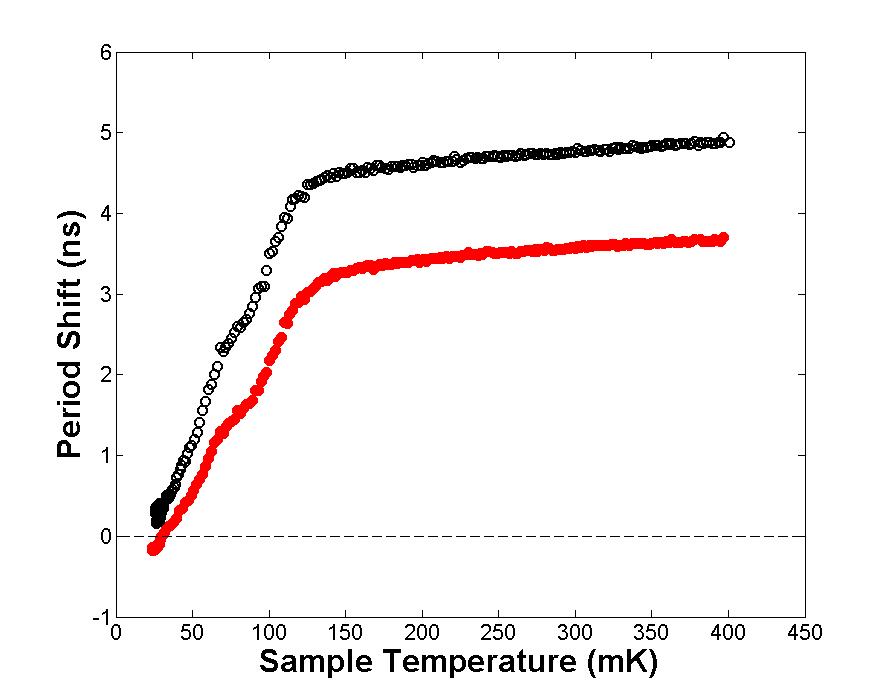}
\caption{Data for the period shift between the period following deformation and the period of the initial sample are plotted as functions of the temperature.  The lower data set, in red, was obtained after a pressure of 10.3 bar had been applied to the flexible diaphragm, while the second set with the larger period increase was obtained after the application of 13.8 bar.}
\label{default}
\end{center}
\end{figure}
These data present a remarkable contrast to expectations based on a scenario of a superfluid-like decoupling of a fraction of the solid moment of inertia. In this scenario, the sample is assumed to be firmly locked to the oscillator above the NCRI onset temperature, and it should be unaffected by the degree of disorder. Thus, the major changes in period, reflecting the growth of the NCRI signal, would occur at temperatures {\it below} the onset, rather than at high temperatures as seen in this experiment. 

\begin{figure}[htbp]
\begin{center}
\includegraphics[width=0.42\textwidth]{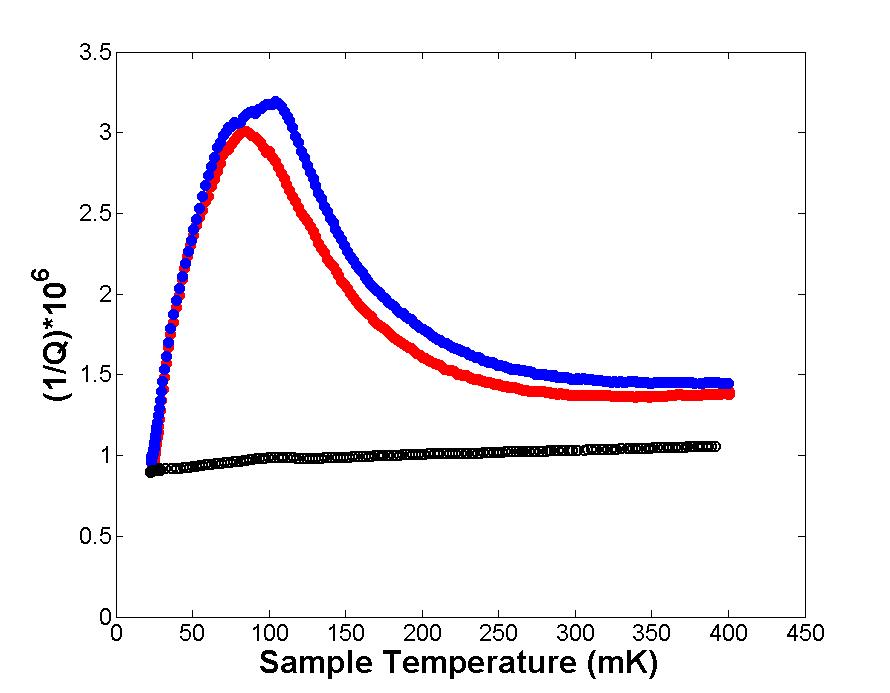}
\caption{Dissipation data for the empty cell are plotted in black, the initial sample in red, and the data for the sample after deformation in blue.  Note that the dissipation data for the empty cell have not been shifted.  At the lowest temperature all three data sets have the same dissipation.}
\label{default}
\end{center}
\end{figure}
The data for the total dissipation, Q$^{-1}$$_{total}$, for the TO corresponding to the period data of Figure 2 are shown in Figure 4 in an unshifted form. The total dissipation can be expressed as the sum of the individual contributions from the empty cell dissipation and dissipation due to the solid, Q$^{-1}$$_{total}$ = Q$^{-1}$$_{empty}$\ $_{cell}$ + Q$^{-1}$$_{solid}$. At the lowest temperature, the dissipation of the empty cell accounts for the entire TO dissipation, with no additional contribution from the solid sample. As the temperature is increased, the solid contribution to the dissipation grows and passes through a peak in the neighborhood of 100 mK.  As the temperature is raised above onset, the solid continues to make a substantial contribution to the total dissipation of the TO above the NCRI onset. Again, this behavior for the dissipation is contrary to what would be expected in the supersolid scenario, where the solid is locked to the TO and should contribute little if any additional dissipation. The dissipative dynamics of the solid $^4$He system have been extensively studied with the TO technique; for a recent work see Hunt et al.$^{16}$

The effects of annealing were investigated in a sample that had been subject to the maximum strain. An initial annealing of this sample took place at a temperature of 0.820 K for a period of 19 hrs. During this time the period decreased 0.6 ns. Following this anneal, the temperature was lowered to 20 mK. The data obtained while warming from this temperature to 1.2 K are shown in Figure 5. 
\begin{figure}[htbp]
\begin{center}
\includegraphics[width=0.42\textwidth]{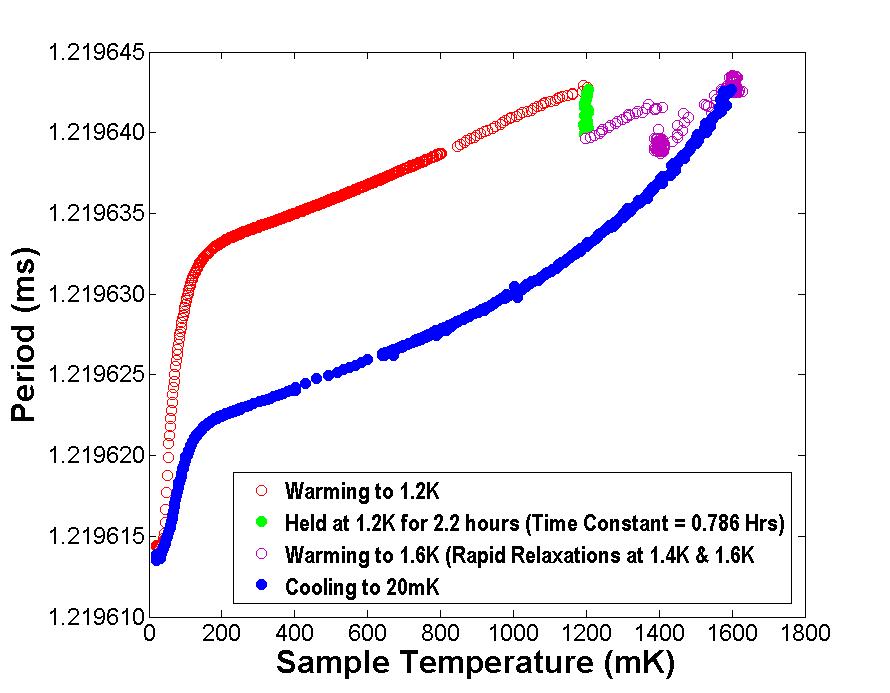}
\caption{Period data for a series of temperature sweeps, including high temperature annealing, are shown as a function of temperature.}
\label{default}
\end{center}
\end{figure}
The sample was further annealed at 1.2 K. Raising the temperature to 1.4 K resulted in an almost discontinuous decrease in the TO period. Finally, the sample was held at 1.6 K for a period of 4 hrs without any further obvious relaxation in the period. The sample was then cooled again to 20 mK. At the lowest temperature the period of the annealed sample returned to within 1 ns of the pre-annealed value. In contrast, at 400 mK the period difference between the pre-annealed sample and the final annealed sample was 11 ns. Although, the dislocation lines introduced at 250 mK by plastic deformation can be readily removed by annealing at temperatures above 1 K, it is also clear that there are other forms of disorder in the sample that are less susceptible to annealing; it is interesting to note, however, that the influence of the remaining unannealed disorder freezes out over much same temperature range, below 0.2 K, as does that of the induced dislocation lines.

The most important conclusions to be drawn from this experiment are that the high temperature period of a TO containing a solid $^4$He sample is strongly dependent on the level of disorder in the sample, with an increase in the TO period following an increase in disorder. The period increase and the additional dissipation associated with sample disorder both vanish at the lowest temperature. An analogous low temperature behavior for the shear modulus has been reported by Day, Syschenko, and Beamish.$^{17}$  They find the low temperature value of the shear modulus to be  relatively unaffected by additional disorder or annealing of the sample; however, several other aspects of their work are not in accord with the present experiment. 

Since the original KC experiments, a number of models based on non-superfluid mechanisms have been developed to explain the NCRI phenomenon.  Among these are the suggestion that a glass transition may take place in the solid,$^{18}$ the two-level system model of Andreev,$^{19}$ as well as a dislocation line model, developed by Iwasa.$^{20}$ 

In the case of an ordinary solid, the sample will be accelerated by an elastic interaction with the walls. When the ratio between the mean radius of the annulus and the radial gap, $\Delta$R/R, is large, as for the present TO, the annular geometry can be approximated by parallel plates separated by $\Delta$R. The displacement of the plates will be R$\theta$$_0$ sin($\omega$,t), where $\theta$$_0$  is the maximum angular displacement of the TO. The motion of the solid is governed by a wave equation, with a solution for the angular displacement of the solid, $\theta$(x,t) = $\theta$$_0$ sin($\omega$t) [cos(2$\pi$$\Delta$Rx/$\lambda$)/cos(¹$\Delta$R/$\lambda$)], where $\lambda$ = (2$\pi$/$\omega$) ($\mu$/$\rho$)$^{1/2}$ is a characteristic wavelength and $\mu$ is an effective shear modulus of the disordered solid. The  x axis, normal to the surface of the plates, has its origin at the midpoint between the plates. The mean angular displacement of the solid, ($\theta$$_0$+$\Delta$$\theta$) sin($\omega$t) = $\theta$$_0$ [$\lambda$/($\pi$$\Delta$R)] tan($\pi$$\Delta$R/$\lambda$)sin($\omega$t) and is slightly larger by the amount, $\Delta$$\theta$, than $\theta$$_0$; ($\Delta$$\theta$/$\theta$$_0$) = [[$\lambda$/($\pi$$\Delta$R)] tan($\pi$$\Delta$R/$\lambda)-1$]. Expanding tan($\pi$$\Delta$R/$\lambda$) to second order in ($\pi$$\Delta$R/$\lambda$), gives ($\Delta$$\theta$/$\theta$$_0$)$\widetilde =$ $\rho$($\Delta$R$^2$$\omega$$^2$/12$\mu$, or $\mu\widetilde=\rho(\Delta R^2\omega^2/(12\Delta\theta/\theta_0$)). The $\omega$$^2$ dependence might be tested with a double frequency TO.

Neglecting frictional terms, the total torque acting on the TO is $\tau_{TO}=-k\theta(t)+\tau_s$= $-$I$_{TO}$ $ \omega^2\theta_0$sin($\omega$t), where $\tau_s$= I$_s$ $\omega^2$($\theta_0+\Delta\theta)$sin$(\omega$t) is the reaction torque due to the solid. I$_{TO}$ and I$_S$ are the moments of inertia of the empty TO and the solid sample, respectively, and k is the torsion constant.  Solving the torque equation for $\omega = 2\pi$/P, where P is the oscillator period, we have P = 2$\pi$$\sqrt{[I_{TO}+I_S(1 + (\Delta\theta/\theta_0))]/k}$.  In the standard supersolid picture, the expression for the TO period is P = 2$\pi$$\sqrt{[I_{TO} +I_S(1 - (\rho_S/\rho))]/k}$. Thus a decrease in ($\Delta$$\theta$/$\theta$$_0$), which might occur as a result of an increase in the shear modulus. will be reflected by a decrease in the TO period that would be interpreted as an increase in ($\rho$$_S$/$\rho$) or the NCRIF in the supersolid scenario.

Iwasa$^{20}$ has suggested that under a shear stress, $\sigma$, there will be an additional displacement strain due to dislocation line motion, $\epsilon=[(\Omega\Lambda)/(\pi\rho \omega{_0}{^2})]\sigma$, where $\Lambda$ is the line density, $\Omega$ is an orientation factor, and $\omega_{0}$ is the resonant frequency for a pinned dislocation line.  This equation serves to define an effective shear modulus $\mu$' = ($\pi\rho \omega{_0}{^2})/(\Omega\Lambda).$   Substituting this value for $\mu$ in the previous expression for ($\Delta\theta/\theta_0$), gives ($\Delta\theta/\theta_0) = (\Omega\Lambda \Delta R^{2}\omega^{2})/(12 \pi\omega{_0}{^2})$. The frequency of pinned lines, $\omega_0$, depends on the length of line between pinning centers that will, in turn, will depend on details of the sample. As an example, taking 5 $\mu$m as the separation between pinning centers, $\omega_{0} = 1.6$ x $10^{7}$ Hz. A 1\% shift in ($\rho_S/\rho$), or ($\Delta\theta/\theta_{0}) = 10^{-2}$ would require $\Omega\Lambda = 2.3$ x $10^{10}$ lines cm$^{-2}$. This value for the line density is several orders of magnitude larger than that observed in ultrasonic experiments.$^{10-13}$ In the Iwasa model, the decrease in period at low temperatures results from additional pinning by $^3$He atoms that condense on the lines as the temperature is lowered. Shorter lines will have higher resonant frequencies and a reduced $\Delta\theta/\theta_{0}$. Above approximately 200 mK, depending on the $^3$He concentration in the sample, the $^3$He atoms will evaporate and the TO period will become relatively independent of temperature. 

Iwasa's model has attractive features, however, the large value required for the line density may be a problem; it is also difficult to see how dislocation line motion might account for the observed NCRI in the case of solid $^4$He contained in porous Vycor glass$^1$ where the line length would be restricted to the pore size of 10 nm.  In addition, a model based solely on dislocation line dynamics does not incorporate the effects of quantum statistics that may be required to explain the absence of an observable NCRI in hcp $^3$He.$^9$

\noindent{\bf Acknowledgements:}  The author thanks Xiao Mi for his assistance, and also thanks J.V. Reppy  and E. N. Smith. He acknowledges profitable discussions with E. Mueller, N. Ashcroft, and V. Ambegaokar, as well as A. Balatsky, M. Graf, and M. Chan. This work has been supported by the National Science Foundation Grant No. DMR-0605864 and the CCMR Grant No. DMR-0520404.

\smallskip
$^{\ast}$Electronic address:  jdr13@cornell.edu
\smallskip

%\begin{list}

\noindent[1] E. Kim and M.H.W. Chan, Nature (London) $\bf 427$, 225 (2004).

\noindent[2]  E. Kim and M.H.W. Chan, Science $\bf 305$, 1941 (2004).

\noindent[3] A.J. Leggett, Phys. Rev. $\bf 25$, 1543 (1970).

\noindent[4]  E. Kim, J.S. Xia, J.T. West, X. Lin, A.C. Clark,and M.H.W. Chan, Phys. Rev. Lett. $\bf 100$, 065301 (2008).

\noindent[5]  A.C. Clark, J.T. West, and M.H.W. Chan, Phys. Rev. Lett. $\bf 99$, 135302 (2007).

\noindent[6]  A.S.C. Rittner and J.D. Reppy, Phys. Rev. Lett. $\bf 97$, 165301 (2006).

\noindent[7]  J. Day and J.R. Beamish, Nature $\bf 450$, 853 (2007).

\noindent[8] A.C. Clark,J.D. Maynard,and M.H.W. Chan,Phys. Rev Lett. B $\bf77$, 155301 (2008).

\noindent[9]  J.T. West, O. Syshchenko, J. Beamish, and M.H.W. Chan, Nature Phys. $\bf 5$, 598 (2009).

\noindent[10]  H. Suzuki, J. Phys. Soc. Japan $\bf 35$, 1472 (1973).

\noindent[11]  D.J. Sanders, H. Kwun, A. Hakata, and C. Elbaum, Phys. Rev. Lett. $\bf 39$, 815 (1977).

\noindent[12]  F. Tsuruoka and Y. Hiki, Phys. Rev. B $\bf 20$, 2702 (1979).

\noindent[13]  I. Iwasa, A. Araki, H. Suzuki, J. Phys. Soc. Japan $\bf 46$, 1119 (1979).

\noindent[14]  G.C. Straty and E.D. Adams, Rev. Sci. Instr. $\bf40$, 1393 (1969).

\noindent[15]  A.S.C. Rittner, W. Choi, E.J. Mueller, and J.D. Reppy, Phys. Rev. B. $\bf80$, 224516 (2009).

\noindent[16]  B. Hunt, E. Pratt, V. Gadagkar, M. Yamashita,A.V. Balatsky, and J.C. Davis, Science $\bf 324$, 632 (2009).

\noindent[17] J. Day, O. Syschenko, and J. Beamish, Phys. Rev. B $\bf79$, 214524 (2009).

\noindent[18]  Z. Nussinov, A.V. Balatsky, M.J. Graf, and S.A. Trugman, Phys. Rev. B $\bf 76$, 014530 (2007).

\noindent[19]  A.F. Andreev, JETP Letters $\bf 85$, 585 (2007).

\noindent[20]  I. Iwasa, Phys. Rev. B $\bf 81$, 104527 (2010).

%\end{thebibliography}

 \end{document}